\documentclass[twocolumn]{aastex631}

\shorttitle{Oxygen-enhanced extremely metal-poor DLAs}
\shortauthors{Welsh et al.}
\graphicspath{{./}{figures/}}
\newcommand{\fe}[1]{Fe\,{\sc ii}~$\lambda#1$}
\newcommand{\oi}[1]{O\,{\sc i}~$\lambda#1$}
\newcommand{\kms}{km\,s$^{-1}$}

\defcitealias{HegerWoosley2010}{HW10}

\begin{document}

\title{Oxygen-enhanced extremely metal-poor DLAs: A signpost of the first stars?}

\correspondingauthor{Louise Welsh}
\email{louise.welsh@unimib.it}

\author[0000-0003-3174-7054]{Louise Welsh}
\affiliation{Centre for Extragalactic Astronomy, Durham University, South Road, Durham DH1 3LE, UK}
\affiliation{Dipartimento di Fisica G. Occhialini, Universit\`a degli Studi di Milano Bicocca, Piazza della Scienza 3, I-20126 Milano, Italy}
\affiliation{INAF – Osservatorio Astronomico di Brera, via Bianchi 46, I-23087 Merate (LC), Italy}

\author[0000-0001-7653-5827]{Ryan Cooke}
\affiliation{Centre for Extragalactic Astronomy, Durham University, South Road, Durham DH1 3LE, UK}

\author[0000-0001-6676-3842]{Michele Fumagalli}
\affiliation{Dipartimento di Fisica G. Occhialini, Universit\`a degli Studi di Milano Bicocca, Piazza della Scienza 3, I-20126 Milano, Italy}

\affiliation{INAF - Osservatorio Astronomico di Trieste, via G. B. Tiepolo 11, I-34143 Trieste, Italy}

\author[0000-0002-5139-4359]{Max Pettini}
\affiliation{Institute of Astronomy, University of Cambridge, Madingley Road, Cambridge CB3 0HA, UK}



\begin{abstract}
We present precise abundance determinations of two near-pristine damped Ly$\alpha$ systems (DLAs) to assess the nature of the [O/Fe] ratio at [Fe/H]~$<-3.0$ (i.e. $<1/1000$ of the solar metallicity). Prior observations indicate that the [O/Fe] ratio is consistent with a constant value, [O/Fe]~$\simeq+0.4$, when $-3<$~[Fe/H]~$<-2$, but this ratio may increase when [Fe/H]~$\lesssim-3$. 
In this paper, we test this picture by reporting new, high-precision [O/Fe] abundances in two of the most metal-poor DLAs currently known. 
We derive values of [O/Fe]~$=+0.50\pm0.10$ and [O/Fe]~$=+0.62\pm 0.05$ for these two $z\simeq3$ near-pristine gas clouds. 
These results strengthen the idea that the [O/Fe] abundances of the most metal-poor DLAs are elevated compared to DLAs with [Fe/H]~$\gtrsim-3$. 
We compare the observed abundance pattern of the latter system to the nucleosynthetic yields of Population III supernovae (SNe), and find that the enrichment can be described by a $(19-25)$~M$_{\odot}$ Population III SN that underwent a $(0.9-2.4)\times10^{51}$~erg explosion. These high-precision measurements showcase the behaviour of [O/Fe] in the most metal-poor environments.
Future high-precision measurements in new systems will contribute to a firm detection of the relationship between [O/Fe] and [Fe/H]. These data will reveal whether we are witnessing a chemical signature of enrichment from Population III stars and allow us to rule out contamination from Population II stars.
\end{abstract}



\section{Introduction} \label{ofe:sec:intro}
The first stars in the Universe are responsible for producing the first chemical elements heavier than lithium. These elements --- known as metals --- irrevocably changed the process of all subsequent star formation and mark the onset of complex chemical evolution within our Universe. Since no metal-free stars have been detected thus far, we know very little about their properties (e.g. their mass distribution) and the relative quantities of the metals that they produced. When the first stars ended their lives, some as supernovae (SNe), they released the first metals into their surrounding environment. The stars that formed in the wake of these (Population III) SNe were born with the chemical fingerprint of the first stars. By studying the chemistry of these relic objects, we can investigate the metals produced by the first stars and, ultimately, trace the evolution of metals across cosmic time. \\
\indent Historically, the fingerprints of the first stars have been studied in the atmospheres of low mass, Population II stars that are still alive today (e.g. \citealt{Cayrel2004, Frebel2005, Aoki2006, Frebel2015, Ishigaki2018, Ezzeddine2019}); the composition of the stellar atmosphere is studied in absorption against the light of the star. 
This process of stellar archaeology, along with simulations of stellar evolution, allows us to infer the elements produced by the first SNe and, subsequently, infer their properties \citep[such as mass, rotation rate, explosion energy, and even the geometry of stellar outflows;][]{WoosleyWeaver1995, ChieffiLimongi2004, Meynet2006,  Tominaga2007, Ekstrom2008, HegerWoosley2010, LimongiChieffi2012}. \\
\indent Extragalactic gas, often seen as absorption along the line-of-sight towards unrelated background quasars, offers a complementary opportunity to study chemical evolution and the first stars \citep{Pettini2008, Penprase2010, Becker2011}. The extragalactic gas clouds that have been studied in absorption to date cover a broad range of metallicity, which appears to increase over time \citep{Rafelski2012, Rafelski2014, Jorgenson2013, Lehner2016, Quiret2016, Lehner2019}. Those whose relative iron abundance is 1/1000 of the solar value (i.e. [Fe/H]~$<-3.0$)\footnote{Here, and throughout this paper,  [X/Y] denotes the logarithmic number abundance ratio of elements X and Y relative to their solar values X$_{\odot}$ and Y$_{\odot}$, i.e. $[{\rm X / Y}] =  \log_{10}( N_{{\rm X}}/N_{{\rm Y}}) - \log_{10} (N_{{\rm X}}/N_{{\rm Y}})_{\odot}$.} are classified as  extremely metal-poor (EMP). These environments have necessarily experienced minimal processing through stars and are therefore an ideal environment to search for the chemical signature of the first stars. \\
\indent Among the least polluted environments currently known, there are three absorption line systems at $z\sim 3-4$ that appear to be entirely untouched by the process of star formation, with metallicity limits of ${\rm [M/H]}\lesssim-4.0$ \citep{Fumagalli2012, Robert2019}; all three are Lyman limit systems (LLSs) whose neutral hydrogen column density is $16.2< \log_{10} N($H\,{\sc i}$)/ {\rm cm}^{-2}<19.0$. These pristine LLSs are a rarity. It is more common to detect absorption line systems that are, at least, minimally enriched with metals. \citet{Crighton2016} reported the detection of a LLS at $z\approx3.5$ with a metal abundance $Z/Z_{\odot} = 10^{-3.4 \pm 0.26}$ whose [C/Si] abundance is consistent with enrichment by either a Population III or Population II star. In order to distinguish between these scenarios, additional metal abundance determinations are required. Distinguishing between the gaseous systems enriched by Population III stars and later stellar populations would allow us to trace the metals produced by the first stars and determine the typical Population III properties. Furthermore, such an investigation will reveal the timescale over which these gas clouds are enriched by subsequent stellar populations. \\
\indent A prime environment to disentangle these chemical signatures are damped Ly$\alpha$ systems (DLAs; $\log_{10} N({\rm H}$\,{\sc i}$)/{\rm cm}^{-2} > 20.3$; see \citealt{Wolfe2005} for a review). Indeed, the most metal-poor DLAs may have been exclusively enriched by the first generation of metal-free stars \citep{Erni2006, Pettini2008, Penprase2010, Cooke2017, Welsh2019}. These high H\,{\sc i} column density gas clouds are self-shielded from external radiation. 
Thus, the constituent metals reside in a single, dominant, ionization state. This negates the need for ionization corrections and leads to reliable gas-phase abundance determinations. 
These systems are most easily studied in the redshift interval $2<z<3$ when the strongest UV metal absorption features are redshifted into the optical wavelength range. 
Only the most abundant elements are typically observed in EMP DLAs, including the $\alpha$-capture elements (C, O, Si, S), some odd atomic number elements (N, Al), and some iron-peak elements (usually, only Fe). Given that these elements trace various nucleosynthetic pathways, these abundant elements are sufficient to understand the properties of the stars that are responsible for the enrichment of EMP DLAs, and tease out the potential fingerprints of the first stars.
\\
\indent Based on the chemical abundances of EMP stars, we have uncovered some signatures of the first stars, including the enhancement of the lighter atomic number elements relative to the heavier atomic number elements. For example, the observed enhancement of carbon relative to iron in EMP stars with a normal abundance of neutron capture elements (i.e. a `CEMP-no' star) may indicate that these stars contain the metals produced by Population III stars (see \citealt{Beers2005} for a review). 
Reminiscent of this signature in stars,
there is tentative evidence of an enhanced [O/Fe] abundance in the most metal-poor DLAs. Specifically, all DLAs with an iron abundance between $-3.0<{\rm [Fe/H]}<-2.0$ are scattered around an [O/Fe] plateau of [O/Fe]$\simeq+0.4$, while those with [Fe/H] $< -3.0$ exhibit a modestly elevated [O/Fe] abundance. 
The plateau in [O/Fe] observed in DLAs with [Fe/H]~$>-3.0$ suggests that the relatively higher metallicity DLAs were all enriched by a similar population of stars, drawn from the same initial mass function (IMF). Since oxygen is predominantly sourced from the supernovae of massive stars, the apparent `inflection' observed in the EMP regime can be explained by three equally exciting possibilities. Relative to the stars that enriched the DLAs with [Fe/H] $>-$3.0, the stars that enriched the most metal-poor DLAs were either: (1) Drawn from an IMF that was more bottom-light; (2) ejected less Fe-peak elements; or (3) released less energy during the explosion that ended their life. All three of these alternatives are signatures of enrichment by a generation of metal-free stars (e.g. \citealt{HegerWoosley2010}). However, the errors associated with the currently available data are too large to confirm this trend. \\
\begin{table*}
\centering
\caption{Journal of observations. \label{ofe:tab:sum}}
\begin{tabular}{llllllllll}
\hline
QS0 & r & $z_{\rm em}$ & $z_{\rm abs}$ & Telescope/ & Wavelength & Resolution & Integration & S/N$^{b}$ & Programme \\
 & (mag) &  &  & instrument & range$^{a}$ (\AA) & (km\,s$^{-1}$) & time (s) &  & ID \\
 \hline
J0955+4116 & 19.38                                              & 3.420 & 3.280 & KECK/HIRES                                                      & 3839 -- 6718                                                   & 6.3                                                        &                                                      32\,400           &   20  &     N162Hb, C320Hb \\
           &                                                    &       &        & KECK/HIRES                                                      & 3242 -- 7805                                                     & 8.3                                                       &                                              14\,400                    &  18 &    A152Hb       \\
J1001+0343 & 17.72                                              & 3.198 & 3.078 & VLT/UVES                                                        & 3282 -- 6652                                                     & 7.3                                                        & 33\,700                                                          &  28   & 083.A-0042(A) \\
           &                                                    &       &        & VLT/UVES                                                        & 3756 -- 10429                                                     & 7.3                                                        & 24\,000                                                          & 11    & 105.20L3.001  \\ \hline
\end{tabular}
\begin{tabular}{l}
     $^{a}$ With some wavelength gaps. \\
     $^{b}$ Near the accessible O\,{\sc i} line --- either $\lambda1302$ or $\lambda1039$
\end{tabular}
\end{table*}
\indent In this paper, we present the detailed chemical abundances of two chemically near-pristine DLAs to study the behaviour of [O/Fe] at the lowest metallicities. These DLAs are found along the line-of-sight to the quasars SDSS J095542.12+411655.3 (hereafter J0955+4116) and SDSS J100151.38+034333.9 (hereafter J1001+0343). 
Previous observations of these quasars have shown that these two gas clouds are among the most metal-poor DLAs currently known. These gas clouds are therefore ideally placed to assess the [O/Fe] inflection in near-pristine environments. 
This paper is organised as follows. Section \ref{ofe:sec:obs} describes our observations and data reduction. In Section \ref{ofe:sec:ana}, we present our data and determine the chemical composition of the two DLAs. We discuss the chemical enrichment histories of these systems in Section \ref{ofe:sec:disc}, before drawing overall conclusions and suggesting future work in Section \ref{ofe:sec:conc}.
\section{Observations} 
\label{ofe:sec:obs}
The data analysed in this paper are either the first high-resolution observations of the DLA (as is the case for J0955+4116), or, we have obtained additional high-resolution observations that target previously unobserved metal lines (as is the case for J1001+0343).\\
\indent The DLA identified along the line of sight to the $m_{r}\approx19.38$ quasar --- J0955+4116 --- was identified by \citet{Penprase2010} as an EMP DLA, based on observations with the Keck Echellete Spectrograph and Imager. We then observed this quasar using the High Resolution Echelle Spectrometer (HIRES; \citealt{Vogt1994}) on the Keck I telescope. We utilised the C1 (7.0 $\times$ 0.861 arcsec slit) and C5 (7.0 $\times$ 1.148 arcsec slit) decker, resulting in a spectral resolution of 49\,000 and 37\,000, respectively. These observations consist of $9\times 3600$\,s exposures using the C1 setup and $4\times 3600$\,s exposures using the C5 setup.\\
\indent Prior observations of the $m_{r}\approx17.7$ quasar --- J1001+0343 --- using the Ultraviolet and Visual Echelle Spectrograph (UVES; \citealt{Dekker2000}) at the European Southern Observatory (ESO) Very Large Telescope (VLT) revealed that the intervening DLA at $z_{\rm abs}\approx 3.078$ is one of the least polluted gas reservoirs currently known. The 9.3 hours of VLT/UVES data presented in \cite{Cooke2011b} indicate that [Fe/H]~$=-3.18\pm 0.15$ and [O/Fe]~$=+0.53\pm0.16$. Given that the typical [O/Fe] abundance observed amongst very metal-poor (VMP; [Fe/H]~$<-2$)  DLAs is $\sim+0.4$, 
the EMP DLA towards J1001+0343 is ideally placed to investigate if the [O/Fe] abundance is elevated at the lowest metallicities. The observations carried out by \cite{Cooke2011b} covered the wavelength range $3282-6652$~\AA. Thus, the iron abundance was determined from observations of the \fe{1608} line. We have secured further observations that focus on red wavelengths and target the stronger \fe{2344} and $\lambda2382$ features of this DLA. The new data on J1001+0343 were collected with UVES ($R~\simeq 40\,000$) throughout the observing period P106 and P108 spanning the wavelength range $3756 - 4985$~\AA\,  and $6705- 10429$~\AA\, using a 0.8 arcsec slit width. 
We acquired $8\times3000$~s exposures on target using $2\times 2$ binning in slow readout mode. A summary of our observations can be found in Table~\ref{ofe:tab:sum}. 

\subsection{Data reduction}
The HIRES data were reduced with the {\sc makee} reduction pipeline while the ESO data were reduced with the EsoRex reduction pipeline. 
Both pipelines include the standard reduction steps of subtracting the detector bias, locating and tracing the echelle orders, flat-fielding, sky subtraction, optimally extracting the 1D spectrum, and performing a wavelength calibration. The data were converted to a vacuum and heliocentric reference frame.

Finally, we combined the individual exposures of each DLA using {\sc uves\textunderscore popler}\footnote{{\sc uves\textunderscore popler} is available from: \\\url{https://github.com/MTMurphy77/UVES\textunderscore popler}}. This corrects for the blaze profile, and allowed us to manually mask cosmic rays and minor defects from the combined spectrum. 
When combining these data we adopt a pixel sampling of 2.5~km\,s$^{-1}$. Due to the different resolutions of the C1 and C5 HIRES deckers, we separately combine and analyse the data collected using each setup. For the regions of the J1001+0343 spectrum $\sim9000$~\AA, that are imprinted with absorption features due to atmospheric $\rm H_{2}O$, we also perform a telluric correction with reference to a telluric standard star. We test the robustness of this correction by also analysing the extracted spectra of the individual exposures (as discussed further in Section~\ref{ofe:sec:j1001}). 
\section{Analysis} 

\label{ofe:sec:ana}
\indent Using the Absorption LIne Software ({\sc alis}) package\footnote{{\sc alis} is available from:\\ \url{https://github.com/rcooke-ast/ALIS}.}--- which uses a $\chi$-squared minimisation procedure to find the model parameters that best describe the input data --- we simultaneously analyse the full complement of high S/N and high spectral resolution data currently available for each DLA. We model the absorption lines with a Voigt profile, which consists of three free parameters: a column density, a redshift, and a line broadening. We assume that all lines of comparable ionization level have the same redshift, and any absorption lines that are produced by the same ion all have the same column density and total broadening. The total broadening of the lines includes a contribution from both turbulent and thermal broadening. The turbulent broadening is assumed to be the same for all absorption features, while the thermal broadening depends inversely on the square root of the ion mass; thus, heavy elements (e.g. Fe) will exhibit absorption profiles that are intrinsically narrower than the profiles of lighter elements, (e.g. C). There is an additional contribution to the line broadening due to the instrument.  For the HIRES and UVES data, the nominal instrument resolutions are $v_{\rm FWHM}= 6.28$~km~s$^{-1}$ (HIRES C1), $v_{\rm FWHM}= 8.33$~km~s$^{-1}$ (HIRES C5), and $v_{\rm FWHM} =7.3$~km~s$^{-1}$ (UVES).
Finally, we note that we simultaneously fit the absorption and quasar continuum of the data. We model the continuum around every absorption line as a low-order Legendre polynomial (typically of order 3). We assume that the zero-levels of the sky-subtracted UVES and HIRES data do not depart from zero\footnote{We visually inspected the troughs of saturated absorption features to confirm this is the case.}. 
In the following sections we discuss the profile fitting for each DLA in turn. 
\begin{figure*}
    \centering
    \includegraphics[width=0.9\textwidth]{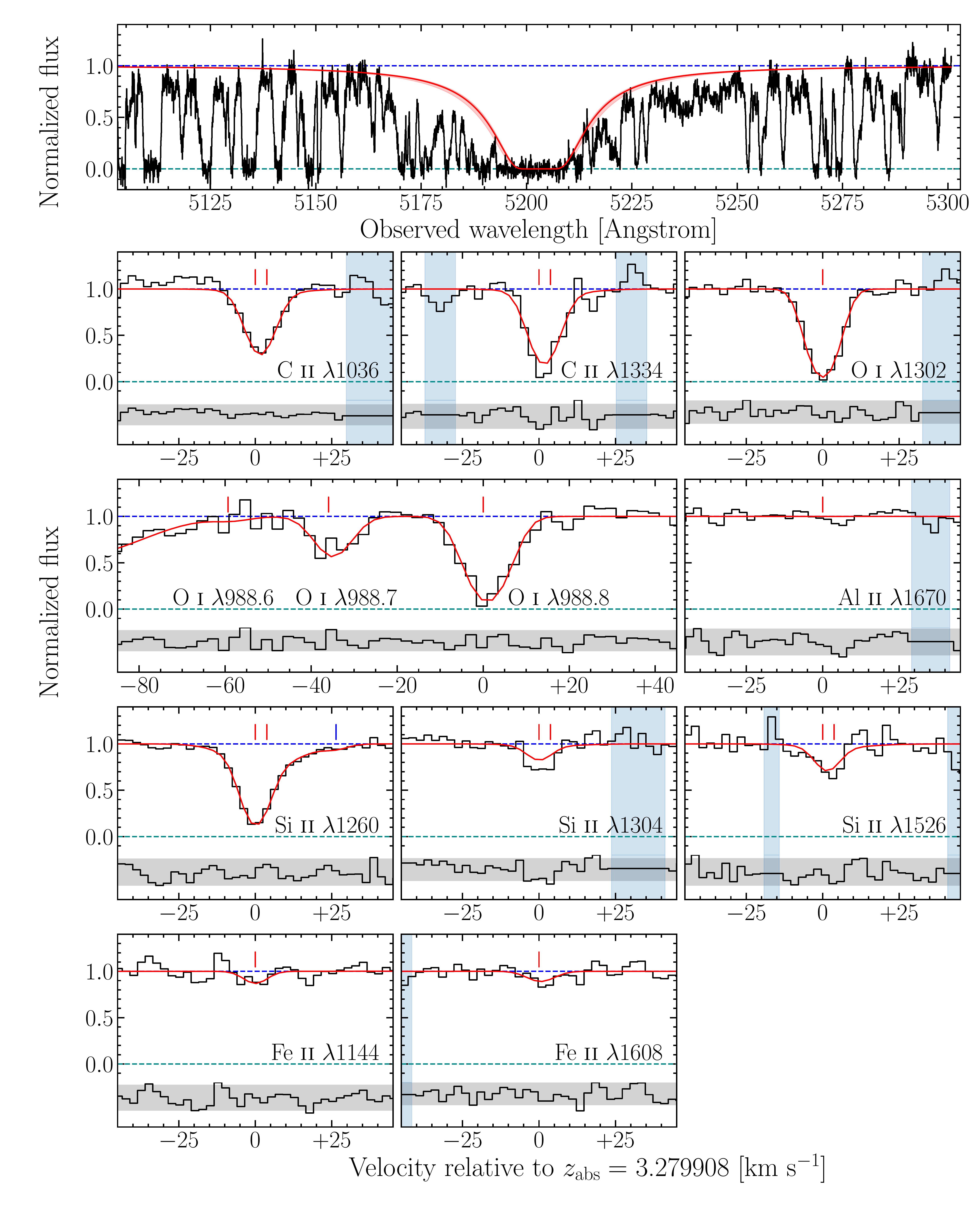}
    \caption{Continuum normalised HIRES data (black histograms) of the absorption features produced by metal ions associated with the DLA at $z_{\rm abs}=3.279908$ towards the quasar J0955$+$4116. The best-fitting model is shown with the red curves. In the top panel, the red shaded band highlights range of H model profiles included in the 1$\sigma$ error bounds. The blue dashed line indicates the position of the continuum while the green dashed line indicates the zero-level. The red ticks above the absorption features indicate the centre of the Voigt line profiles. In the left panel of the fourth row, the blue tick indicates the centre of the Fe\,{\sc ii} 1260 feature. Below the zero-level, we show the residuals of this fit (black histogram) where the grey shaded band encompasses the $2\sigma$ deviates between the model and the data. The vertical blue shaded bands indicate the regions of the spectrum not included in the fit.}
    \label{fig:j0955}
\end{figure*}
{\footnotesize \begin{table}
	\centering
	\caption{Ion column densities of the DLA at $z_{\rm abs}=3.279908$ towards the quasar J0955$+$4116. The quoted column density errors are the $1\sigma$ confidence limits.}
	\label{tab:j0955}
	\tabcolsep=0.10cm
\begin{tabular}{lcccc}
\hline
Ion        & Transitions & log$_{10}$ $N$(X)/cm$^{-2}$  & [X/H] & [X/Fe] \\ 
  & used [\AA] &  &  &  \\ \hline \hline
H\,{\sc i}   & 1215             & 20.21 $\pm$ 0.05   &  ---             &  ---   \\
C\,{\sc ii}  & 1036, 1334       & 13.73 $\pm$ 0.14   & $-3.00 \pm 0.06$ & $-0.05 \pm 0.10$   \\
O\,{\sc i}   & 988, 1302       & 14.45 $\pm$ 0.04   & $-2.45 \pm 0.06$ & $+0.50 \pm 0.10$    \\
Al\,{\sc ii} & 1670             & $\leq11.25^{\rm a}$   & $\leq -3.41$ &  $\leq-0.46 $  \\
Si\,{\sc ii} & 1260, 1304,  & 12.89 $\pm$ 0.06  & $-2.94 \pm 0.06$ &  $+0.01 \pm 0.09$  \\
  & 1526 &  &  &  \\ 
Fe\,{\sc ii} &   1144, 1260,    & 12.73 $\pm$ 0.09   & $-2.95 \pm 0.10$ &  ---                 \\ 
  & 1608 &  &  &  \\ 
\hline
\end{tabular}
\begin{tabular}{l}
     $^{a}$ $3\sigma$ upper limit on column density. \\
\end{tabular}
\end{table}}
\subsection{J0955+4116}
J0955+4116 is best modelled with two gaseous components at $z_{\rm abs}= 3.279908 \pm 0.000002$ and $z_{\rm abs}=3.27996 \pm 0.00001$ ($\Delta v = 4 \pm 1$~\kms) for all singly ionized species (except Fe\,{\sc ii}), and just the former component for neutral species (i.e. O\,{\sc i}). We assume the temperature is $T=1\times10^{4}$~K  (a value that is typical for a metal-poor DLA; see \citealt{CookePettiniJorgenson2015}, \citealt{Welsh2020}, \citealt{Noterdaeme2021}) and find that the turbulent components are $b=3.3\pm 0.2$~km\,s$^{-1}$ and $b=14.2\pm1.5$~km\,s$^{-1}$ respectively. The data, along with the best-fitting model are presented in Figure \ref{fig:j0955}, while the corresponding column densities are listed in Table \ref{tab:j0955}. These results are unchanged when the assumed temperature varies between $T \sim(0.5 - 1.2)\times 10^{4}$~K --- the range of values that have been measured in other metal-poor DLAs \citep{CookePettiniJorgenson2015}. We find the data are best modelled when we allow for small velocity offsets between the features observed using the C1 and C5 deckers. These are found to be $\lesssim 1.5$~\kms\ and account for potential differences between the wavelength calibrations of the data taken with the two HIRES setups.

Note, we only use the neutral component (identified in the O\,{\sc i} absorption) to infer the relative chemical abundances of this gas cloud; the component at $z_{\rm abs}=3.27996$ likely arises from ionized gas as indicated by the lack of concurrent absorption from any neutral species. The neutral component at $z_{\rm abs}=3.279908$ constitutes $\sim70\%$ to the total absorption in Si\,{\sc ii}. We find that [Fe/H]~$=-2.95\pm 0.10$ while [O/Fe]~$=+0.50\pm0.10$. This places the DLA towards J0955+4116 at the cusp of the EMP regime where the plateau in [O/Fe] may change. Here, and subsequently, the errors are given by the square root of the diagonal term of the covariance matrix calculated by {\sc alis} at the end of the fitting procedure. We note that, while the individual Fe\,{\sc ii} features may be relatively weak, the simultaneous analysis of the full complement of data results in a 4.8$\sigma$ detection of Fe. 
\subsection{J1001+0343}
\label{ofe:sec:j1001}
\begin{figure*}
    \centering
    \includegraphics[width=0.8\textwidth]{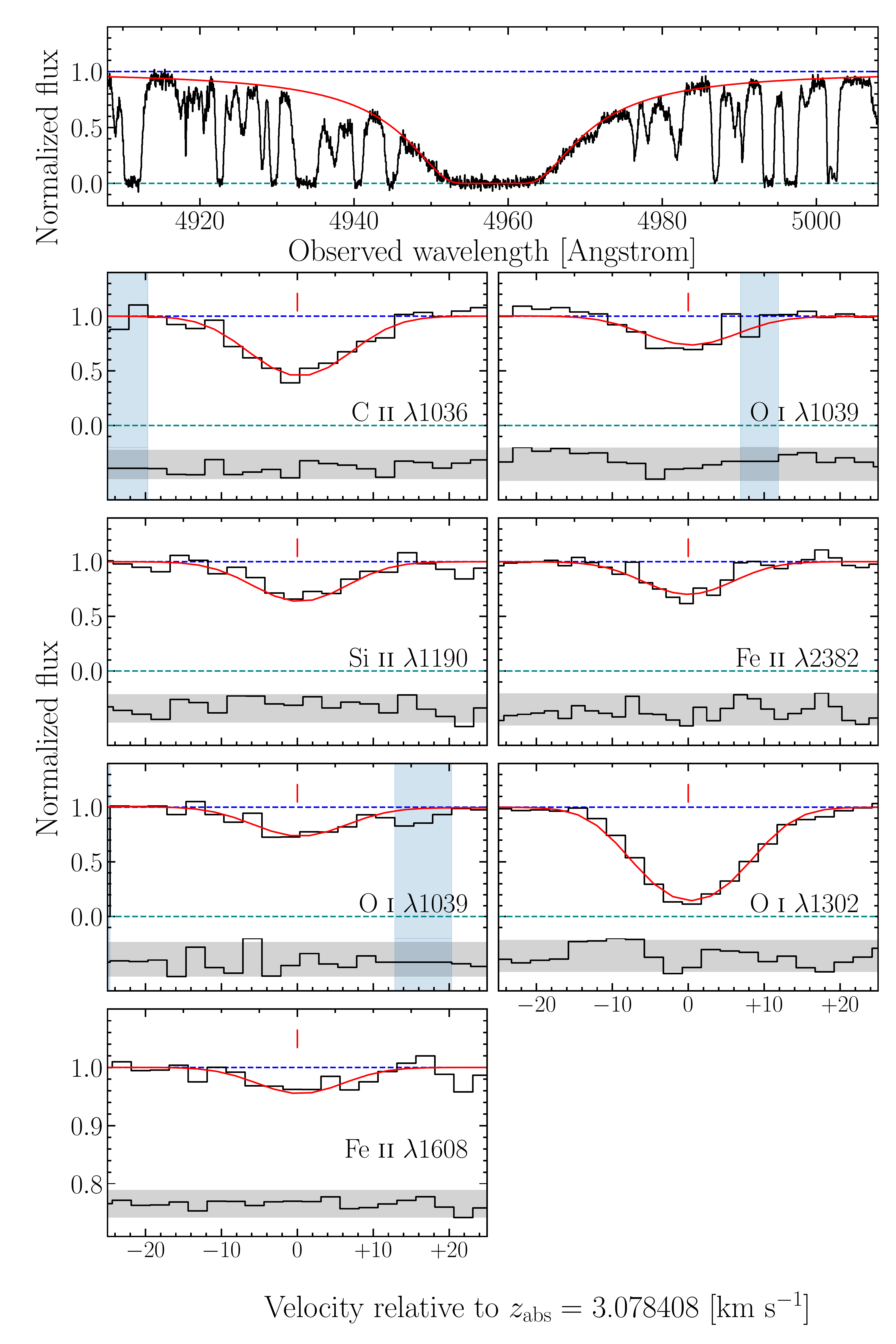}
    \caption{Continuum normalised UVES data (black histograms) of the absorption features produced by metal ions associated with the DLA towards J1001$+$0343. The top row shows H\,{\sc i}. The second and third rows show our new programme data (programme ID: 105.20L3.001) while the bottom two rows show the archival data (programme ID: 083.A-0042(A)). The best-fitting model profiles are shown as red curves. The blue dashed line indicates the position of the continuum while the green dashed line indicates the zero-level. The red ticks above the absorption features indicate the centre of the Voigt line profiles. Below the zero-level, we show the residuals of this fit (black histogram) where the grey shaded band encompasses the $2\sigma$ deviates between the model and the data. The vertical blue shaded bands indicate the regions of the spectrum not included in the fit.}
    \label{fig:j1001}
\end{figure*}

\begin{figure*}
    \centering
    \includegraphics[height=0.85\textheight]{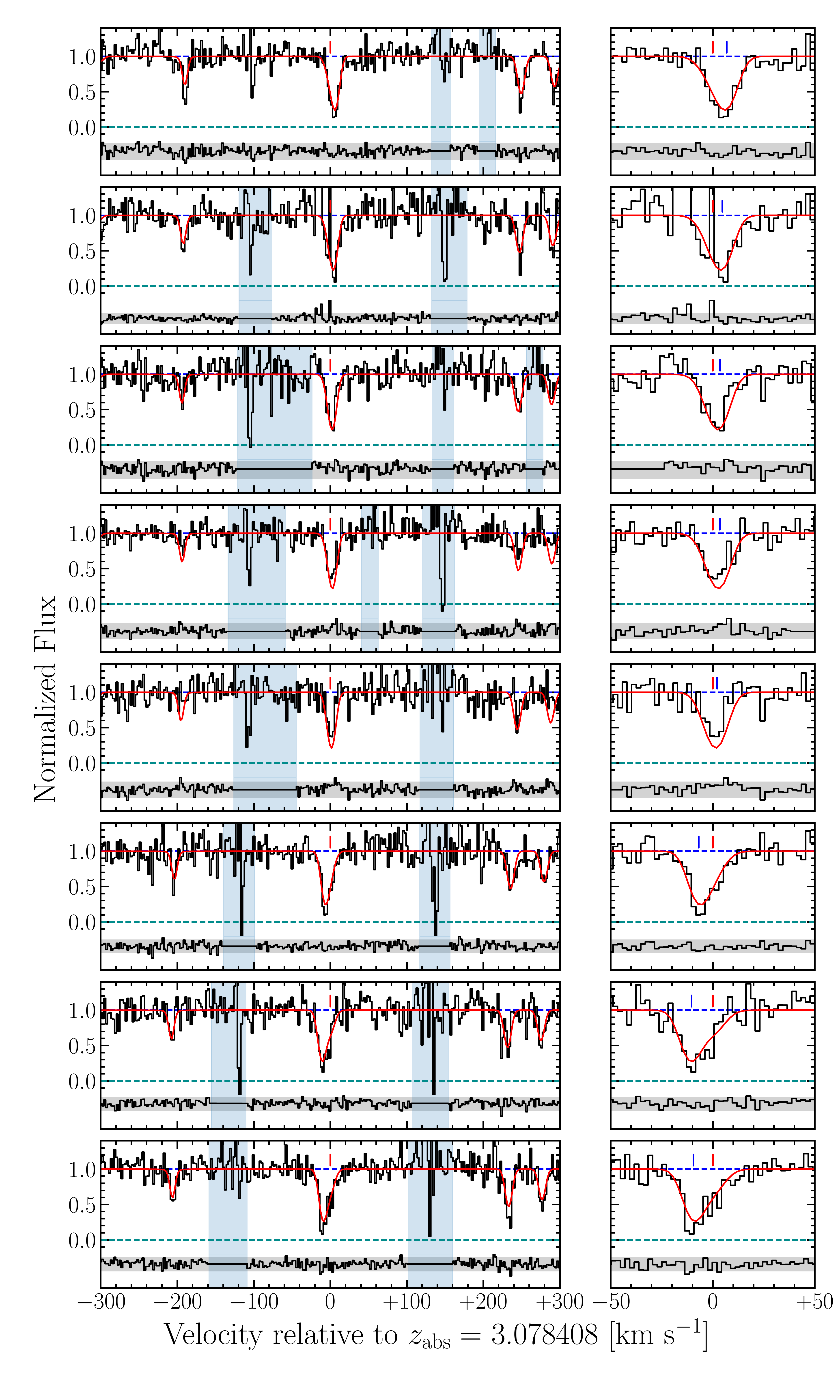}
    \vspace{-0.5cm}
    \caption{Continuum normalised UVES data (black histograms) of the absorption features produced by \fe{2382} associated with the DLA towards J1001$+$0343 along with the surrounding telluric absorption features for each exposure. The right panels show a zoom in of the left panels. Overplotted in red is our best-fitting model of the \fe{2382} feature and the telluric absorption. The blue dashed line indicates the position of the continuum while the green dashed line indicates the zero-level. The red ticks above the absorption features indicate the centre of the Voigt line profiles of the DLA, while the blue tick marks indicate the line centre of the telluric features. Below the zero-level, we show the residuals of this fit (black histogram) where the grey shaded band encompasses the $2\sigma$ deviates between the model and the data. The vertical blue shaded bands indicate the regions of the spectrum not included in the fit.}
    \label{fig:fe}
\end{figure*}

{\footnotesize \begin{table}
	\centering
	\caption{Ion column densities of the DLA at $z_{\rm abs}=3.078408$ towards the quasar J1001$+$0343. The quoted column density errors are the $1\sigma$ confidence limits.}
	\label{tab:j1001}
	\tabcolsep=0.10cm
\begin{tabular}{lcccc}
\hline
Ion  & Transitions & log$_{10}$ $N$(X)/cm$^{-2}$ & [X/H] & [X/Fe] \\
  & used [\AA] &  &  &  \\ \hline \hline
H\,{\sc i}   &  1215                   & 20.20 $\pm$ 0.05 &      ---         &  ---   \\
C\,{\sc ii}  &  1036, 1334             & 13.57 $\pm$ 0.01  & $-3.06 \pm 0.05$ &  $+0.19 \pm 0.05$   \\
N\,{\sc i}   &  1200                   & $\leq12.50^{\rm a}$ & $\leq-3.53$ &  $\leq-0.28$ \\
O\,{\sc i}   &  1039, 1302             & 14.26 $\pm$ 0.01  & $-2.63 \pm 0.05$ &  $+0.62 \pm 0.05$   \\
Si\,{\sc ii} &  1190, 1260, & 12.86 $\pm$ 0.01  & $-2.85 \pm 0.05$ &  $+0.40 \pm 0.05$   \\
  & 1304, 1526 &  &  &  \\
S\,{\sc ii}   &  1253                   & $\leq12.91^{\rm a}$ & $\leq-2.41$ &  $\leq+0.84$ \\
Fe\,{\sc ii} &  1608, 2382       & 12.42 $\pm$ 0.05  & $-3.25 \pm 0.07$ &      ---            \\ \hline
\end{tabular}
\begin{tabular}{l}
     $^{a}$ $3\sigma$ upper limit on column density. \\
\end{tabular}
\end{table} }
J1001+0343 is best modelled with one component at $z_{\rm abs}=3.078408 \pm 0.000006$ with a  turbulence \\ $b=6.3 \pm 0.4$~km\,s$^{-1}$ and temperature $T=(1.0 \pm 0.6)\times 10^{4}$~K. The 
data, along with the best-fitting model are presented in Figure \ref{fig:j1001}, while the corresponding column densities are listed in Table \ref{tab:j1001}. We find that [Fe/H]~$=-3.25\pm0.07$ and [O/Fe]~$=+0.62 \pm 0.05$. All reported column densities are consistent with the previous determinations by \citet{Cooke2011b}, but with a reduced error; in particular, the new data reported here have allowed the precision on the [O/Fe] measurement to be improved by a factor of three, from 0.15 to 0.05. To ensure the errors associated with these abundance determinations are robust, we have performed some additional checks. First, near the absorption features of interest, we have ensured that the fluctuations in the continuum are well-described by the error spectrum in this region. This ensures we are not underestimating the error associated with the data. Second, we have refit the O\,{\sc i} and Fe\,{\sc ii} features using a Monte Carlo approach, described in \citet{Fossati2019}, and converged on abundance determinations that are consistent within $1\sigma$. \\
\indent When analysing the DLA towards J1001+0343, we adopt two approaches for modelling the Fe\,{\sc ii} absorption features. The \fe{2344} and $\lambda2382$ features fall in regions of the spectrum that are impacted by telluric absorption; the DLA absorption features are therefore partially blended with telluric features to varying degrees of severity. Prior to combining the individual DLA exposures, we remove these features using the spectrum of a telluric standard star. The resulting data near the \fe{2382} line, after performing this correction, are shown in the right panel of the third row of Figure~\ref{fig:j1001}. To ensure that the telluric correction has not introduced any artefacts in the data, we simultaneously fit the standard star spectrum and all of the individual quasar exposures (uncorrected for telluric absorption). The results of this fitting procedure are shown in Figure~\ref{fig:fe}. From this figure it is clear that, while the \fe{2382} feature is partially blended with a telluric absorption line, the range of dates used to observe this target results in a sequential shift in the position of the telluric feature relative to the Fe line of interest. In the top right panel of Figure~\ref{fig:fe}, the telluric absorption is $\sim+5$~\kms\ from the \fe{2382} line center (as indicated by the blue tick mark). In the bottom right panel, the telluric absorption is $\sim-10$~\kms\ from the \fe{2382} line center. When jointly analysing these data, this shift allows us to capture an accurate profile of both the telluric and Fe\,{\sc ii} features. Note, the centroid of the \fe{2382} line is tied to the other DLA absorption features, while the centroid of the telluric feature is fixed from other telluric lines in the standard star spectrum. Using this approach we find a total Fe\,{\sc ii} column density consistent with our analysis of the corrected combined spectrum. The value we report in this paper is based on the fits to the individual exposures.

These new data confirm that the DLA towards J1001+0343 is one of the most iron-poor DLAs currently known. The new found precision afforded for the iron column density allow us to conclude that [O/Fe] is significantly elevated in this DLA compared to the plateau observed at higher metallicity. Before discussing the origin of this elevation, we perform some simulations to support our analysis.
\begin{figure*}
    \centering
    \includegraphics[width=\textwidth]{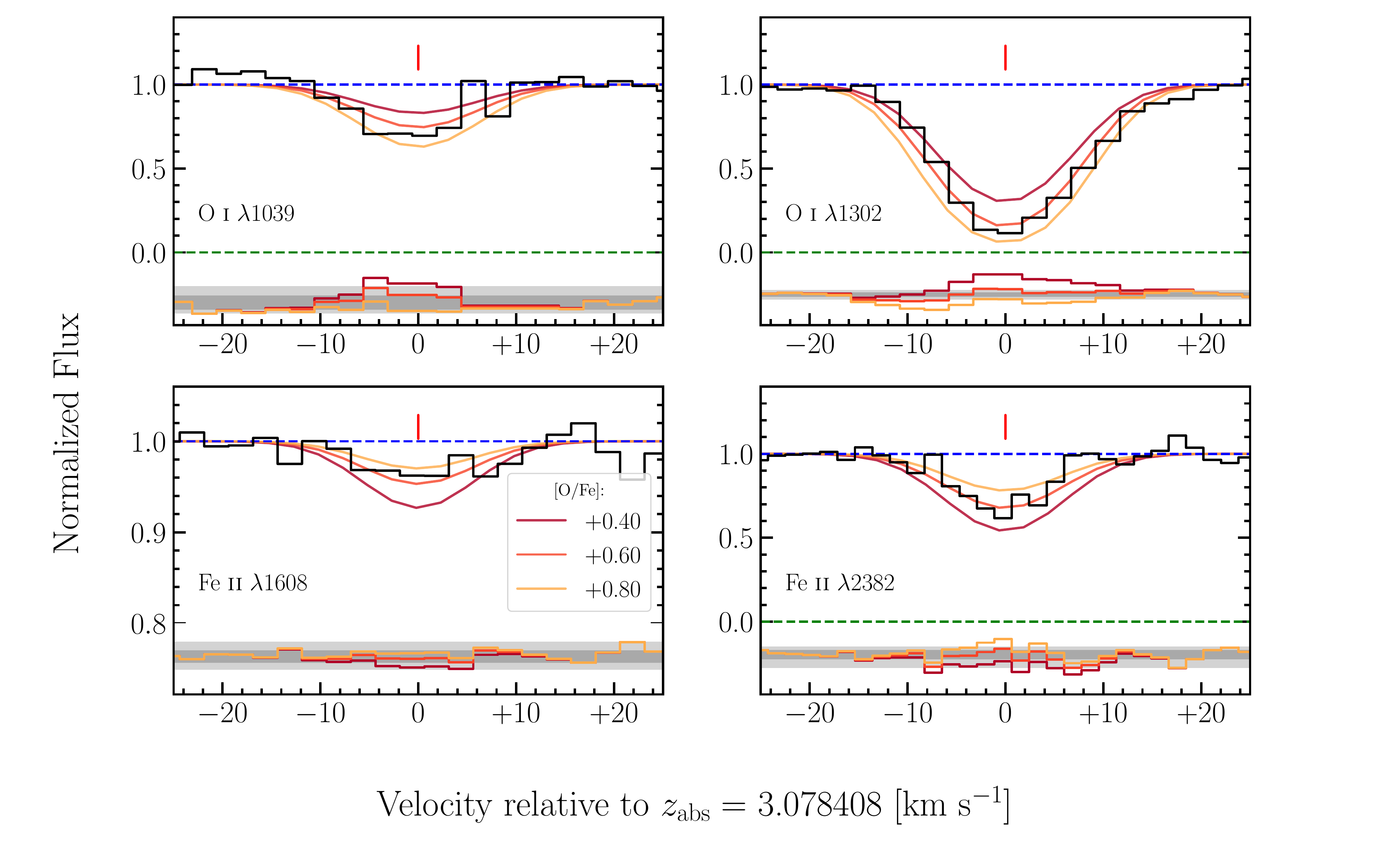}
    \caption{Models of O\,{\sc i} and Fe\,{\sc ii} absorption components overplotted on the UVES data of J1001+0343. The colour indicates the underlying abundance ratio [O/Fe] as highlighted in the legend. The top panels shows the O\,{\sc i} data and the corresponding O\,{\sc i} model profiles assuming a fixed Fe\,{\sc ii} column density. The bottom panels show the corresponding Fe\,{\sc ii} model profiles assuming a fixed O\,{\sc i} column density. These fixed values represent the best fit column densities of the data. The blue (green) dashed lines represents the continuum (zero) levels of the data. Note the different  y-axis range of the \fe{1608} panel. Below each spectrum, we show the residuals of this fit (colored histograms) where the light (dark) grey shaded band encompasses the $2\sigma$ ($1\sigma$) deviates between the model and the data.}
    \label{fig:ofe_mod}
\end{figure*}

\subsection{Mock models}
To further test if the DLA towards J1001+0343 exhibits an elevated [O/Fe] ratio, we simulated the O\,{\sc i} and Fe\,{\sc ii} absorption line profiles that would be expected given different intrinsic [O/Fe] abundance ratios. To achieve this, we take the best fit cloud model from our modelling procedure and generate synthetic model profiles varying the column density of either O\,{\sc i} or Fe\,{\sc ii}. The results of this test are shown in Figure~\ref{fig:ofe_mod}. The top row shows the observed UVES data centered on the \oi{1039} (left) and \oi{1302} (right) absorption features. Overplotted on these data are the model profiles that would be expected if the underlying [O/Fe] ratio were $+0.4$, $+0.6$, and $+0.8$. To generate these profiles, we assume that the Fe\,{\sc ii} column density is fixed (at the value given by our best fit model) and then vary the column density of O\,{\sc i} accordingly. In the bottom row we show the UVES data centered on the \fe{1608} (left) and \fe{2382} (right) lines. The overplotted models show the same underlying [O/Fe] abundance ratios as the top row, however, in this case we vary the column density of Fe\,{\sc ii} (while the column density of O\,{\sc i} is fixed at the value given by our best fit model). Below these data we show the residual fits between the model and the data. The grey shaded band represents the $2\sigma$ limits provided by our best fit model ([O/Fe]~$=+0.62\pm0.05$) and the data. In each panel (except  for the very weak \fe{1608} line), the residual showcasing the fit of the [O/Fe]~$=+0.4$ model (dark red; see legend) is outside of the $2\sigma$ range. From this analysis, we therefore conclude that these data show an [O/Fe] abundance ratio that is inconsistent with the plateau seen at [Fe/H]~$>-3$.

\section{Discussion}
\label{ofe:sec:disc}

\begin{figure*}
    \centering
    \includegraphics[width=\textwidth]{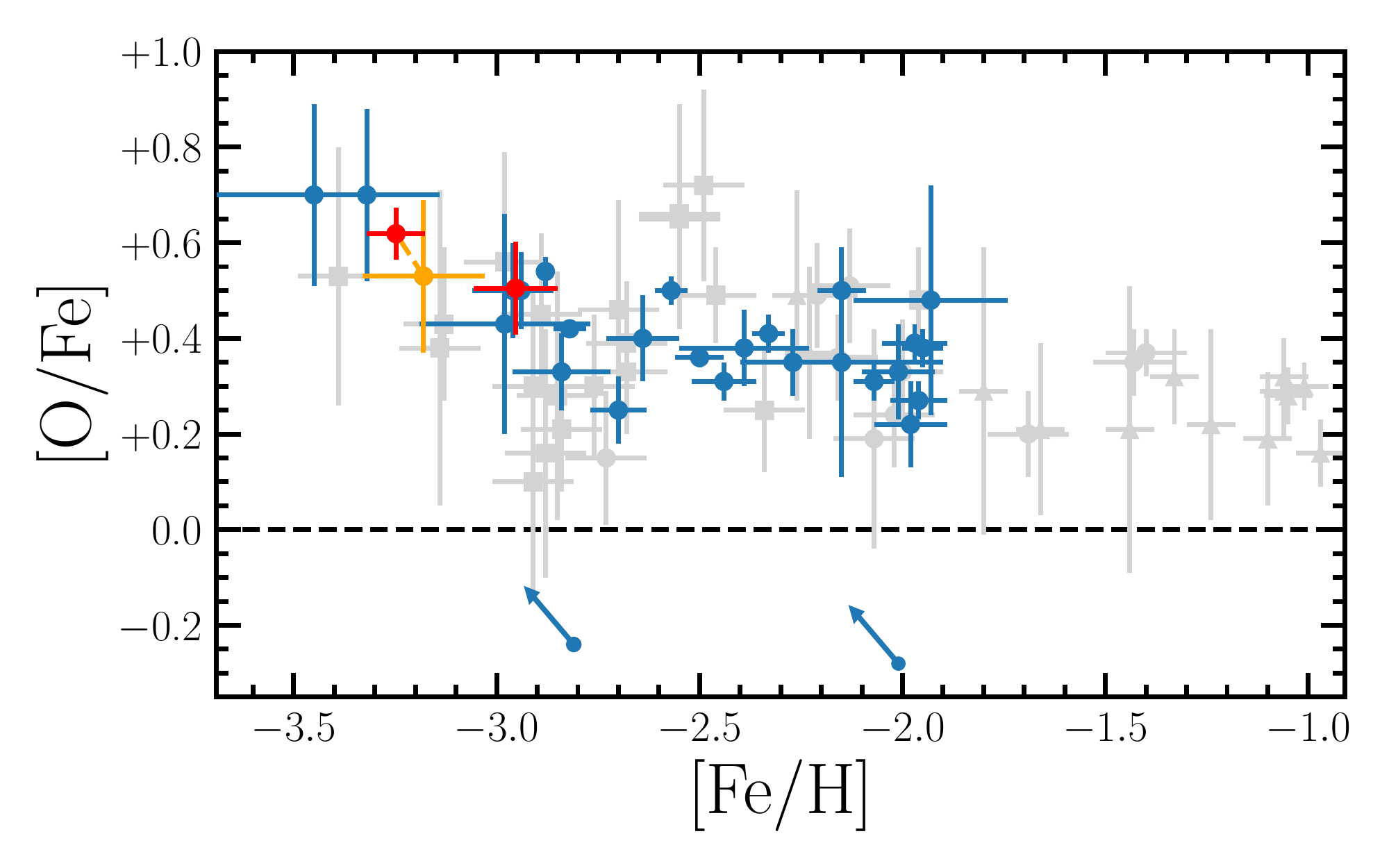}
    \caption{[O/Fe] vs [Fe/H] for all metal-poor DLAs and sub-DLAs (blue circles) plotted alongside [O/Fe] of metal-poor stars (light gray) --- the shape of the marker indicates the source of the stellar data: circles - \cite{Garcia2006}, squares - \cite{Cayrel2004}, triangles - \cite{Nissen2002} (all values are those presented in the \citealt{Cooke2011b} reanalysis). The two [O/Fe] abundances reported here are shown in red. The previous measurement of J1001+0343 is marked in orange and is connected to the latest measurement by a dashed line. The black dashed line indicates the solar relative abundance. }
    \label{fig:ofe_stars}
\end{figure*}

The primary goal of this paper is to assess if EMP DLAs (those with ${\rm [Fe/H]}<-3.0$) exhibit an enhanced [O/Fe] abundance relative to VMP DLAs (those with $-3.0<{\rm [Fe/H]}<-2.0$). 
Figure~\ref{fig:ofe_stars} shows the [O/Fe] abundance ratio as a function of the [Fe/H] metallicity for the DLAs analysed in this work (red symbols), together with literature measurements of VMP and EMP DLAs (blue symbols) and stars (gray symbols). The DLA abundances used to produce this figure are given in Appendix~\ref{ofe:append:data}. 
The new data reported here confirm that the DLA towards J1001+0343 is a bonafide EMP DLA, while the DLA towards J0955+4116 is on the cusp of the EMP regime.
Qualitatively, the [O/Fe] values of these DLAs are consistent with the trend of an increased [O/Fe] ratio below [Fe/H]~$<-3$. Given the metallicity of J0955+4116 ([Fe/H]~$=-2.95\pm 0.10$), it is reasonable to expect that the [O/Fe] abundance ratio of this DLA is consistent with the plateau seen at higher metallicity. The high precision [O/Fe] determination of J1001+0343 has strengthened the evidence that this system exhibits an elevated [O/Fe] ratio. 
With this new measurement, we determine the mean of this relative abundance ratio for the EMP DLAs shown in Figure~\ref{fig:ofe_stars}: $[\langle{\rm O/Fe}\rangle]=+0.67 \pm 0.04$. For VMP DLAs, $[\langle{\rm O/Fe}\rangle]=+0.40 \pm 0.08$; the mean [O/Fe] abundance reported for EMP DLAs is significantly divergent ($3\sigma$) from the mean [O/Fe] abundance reported for VMP DLAs. 
There are no obvious selection biases that may have caused this apparent inflection of [O/Fe] at the lowest metallicities probed. Furthermore, these results are consistent with the recent evaluation of [O/Fe] across EMP stars when the stellar spectra are analysed using 3D non-LTE models \citep{Amarsi2019}. Specifically, their analysis finds [O/Fe]$_{\rm EMP \star}$~$\sim +0.7$. 
If this trend of an elevated [O/Fe] abundance is confirmed with future measurements of EMP DLAs, it will highlight that EMP DLAs exhibit a distinct chemical enrichment relative to VMP DLAs; the source of the elevated [O/Fe] in these metal-poor DLAs may be attributed to enrichment by a generation of Population III stars \citep{HegerWoosley2010}. However, before we consider the abundance pattern of this DLA in relation to the yields of Population III SNe, we first examine possible origins of this elevation.

\begin{figure*}
    \centering
    \includegraphics[width=\textwidth]{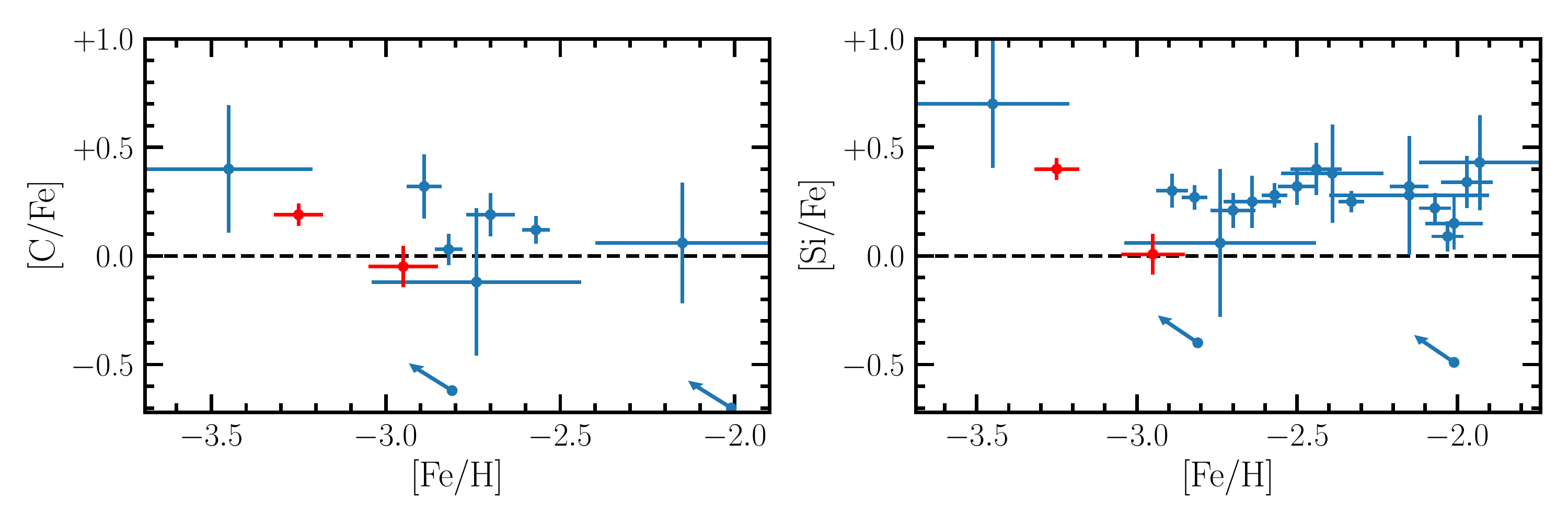}
    \caption{[C/Fe] (left) and [Si/Fe] (right) vs [Fe/H] for all metal-poor DLAs and sub-DLAs (blue circles). The abundances of the DLAs reported here are shown in red. The black dashed line indicates the solar relative abundance.}
    \label{fig:xfe}
\end{figure*}

\subsection{Origin of elevation}
\indent Dust depletion is expected to be minimal in VMP DLAs \citep{Pettini1997,Akerman2005, Vladilo2011, Rafelski2014}. However, if the depletion of metals onto dust grains is unaccounted for, it will lead to artificially low metal abundance determinations for refractory elements. It is therefore useful to rule out its impact in the DLAs presented here. Depletion studies compare the relative abundances of elements in DLAs to the expected nucleosynthetic ratio which can be inferred from the abundances of stars of similar metallicity. O is minimally depleted onto dust grains \citep{SpitzerJenkins1975, Jenkins2009, Jenkins2017}. Both Si and Fe are refractory elements, and are partially depleted onto dust grains but at different rates.
As shown in Figure~\ref{fig:ofe_stars}, both metal-poor halo stars and VMP DLAs exhibit an identical evolution of the [O/Fe] ratio \citep[see also Figure 12 of ][]{Cooke2011b}. 
Given this agreement, we therefore expect dust depletion to be minimal in DLAs that have a metallicity [O/H]~$<-2$. \\
We can also use the [Si/Fe] abundance ratio to explore the possibility of dust depletion. The most metal-poor stars and the most metal-poor DLAs appear to have a metallicity independent evolution of [Si/Fe] when [Fe/H]~$<-2$. For stars, the plateau occurs at [Si/Fe]~$=+0.37\pm 0.15$ \citep{Cayrel2004}. While for DLAs, the plateau occurs at [Si/Fe]~$=+0.32 \pm 0.09$ \citep{Wolfe2005, Cooke2011b}. 
The [Si/Fe] of both J0955+4116 and J1001+0343 are consistent with the plateau seen in metal-poor DLAs (see Figure~\ref{fig:xfe}). We therefore do not expect dust depletion to be the source of the elevated [O/Fe] abundance ratio in the EMP regime. Volatile elements, like S and Zn, are less readily depleted onto dust grains than Si and Fe \citep{Savage1996, Jenkins2009, Jenkins2017}, however these elements are not currently accessible for the EMP DLAs studied here. The advent of the next generation of $30-40$~m telescopes will make these abundance determinations possible for EMP DLAs. \\
\indent We also find it unlikely that the cloud model introduces a systematic [O/Fe] enhancement; the O\,\textsc{i} and Fe\,\textsc{ii} column densities of both DLAs are derived from at least one weak absorption line. Furthermore, we note that ionization effects cannot explain this behaviour at low metallicity; 
the presence of an unresolved component of ionized gas containing Fe\,{\sc ii} would not affect the O\,\textsc{i} column density, but would lead to an overestimate the Fe\,\textsc{ii} column density. Thus, accounting for ionized gas would only act to further \emph{increase} the [O/Fe] ratio.
We therefore conclude that the elevated oxygen to iron abundance ratio observed for the EMP DLA towards J1001+0343 is intrinsic to the DLA.

\subsection{Stochastic enrichment}

In the previous section, we concluded that the observed [O/Fe] ratio is intrinsic to the DLA towards J1001+0343. We now explore the possibility that this DLA has been enriched by the first generation of stars. Specifically, we compare the observed abundance pattern of this DLA to those predicted by a stochastic chemical enrichment model developed in previous work \citep{Welsh2019}. This model describes the underlying mass distribution of the enriching stellar population using a power-law: $\xi (M) = kM^{-\alpha}$, where $k$ is a multiplicative constant that is set by the number of enriching stars that form between a given mass range:
\begin{equation}
\label{eqn:stoc}
    N_{\star} = \int_{M_{\rm min}}^{M_{\rm max}} kM^{-\alpha}{\rm d}M
\end{equation}
Since the first stars are thought to form in small multiples, this underlying mass distribution is necessarily stochastically sampled. We utilise the yields from simulations of stellar evolution to construct the expected distribution of chemical abundances given an underlying IMF model. These distributions can then be used to assess the likelihood of the observed DLA abundances given an enrichment model. \\
\indent In our analysis, we use the relative abundances of [C/O], [Si/O], and [Fe/O] when investigating the enrichment of the DLA towards J1001+0343. 
We compare the measured abundances to the nucleosynthetic yield calculations of massive ($>10$~M$_{\odot}$) metal-free stars from \citet{HegerWoosley2010} (hereafter \citetalias{HegerWoosley2010}). These yields have been calculated as a function of the progenitor star mass ($M$), the explosion energy of their supernova ($E_{\rm exp}$), and the mixing between the different stellar layers ($f_{\rm He}$). The explosion energy is a measure of the final kinetic energy of the ejecta at infinity while the mixing between stellar layers is parameterised as a fraction of the helium core size. For further details, see both \citetalias{HegerWoosley2010} and \citet{Welsh2019}. \\
\indent While the \citetalias{HegerWoosley2010} yields have been calculated for metal-free stars, they are also representative of EMP Population II CCSNe yields (at least for the elements under consideration in this work); this can be seen by comparison with the \citet{WoosleyWeaver1995} yields of metal-enriched massive stars. As a result, in previous studies of near-pristine absorption line systems, it has been difficult to distinguish between enrichment by Population II and Population III stars. Fortunately, in our analysis, we can take advantage of this degeneracy and consider the \citetalias{HegerWoosley2010} yields to be representative of both Population II and Population III SNe yields. \\
\indent The default enrichment model, described above, contains six free parameters ($N_{\star}, M_{\rm min}, M_{\rm max}, \alpha, E_{\rm exp}$, and $f_{\rm He}$). 
We are using three relative abundances to assess the enrichment of the DLA towards J1001+0343. Thus, we cannot simultaneously investigate these six parameters. We can, however, make some simplifications. The underlying IMF of the first stars remains an open question; this is not the case for massive Population II stars. The Population II IMF for stars of mass $M\gtrsim10\,{\rm M}_{\odot}$ is expected to be well-described by a Salpeter IMF (i.e. $\alpha$ = 2.35 in Equation~\ref{eqn:stoc})\footnote{This was the first local measurement of the stellar IMF \citep{Salpeter1955} --- see \citet{Bastain2010} for a review.}. 
Under the assumption of a Salpeter IMF, if the number of enriching stars we derive is large, then this may imply that Population II stars are the dominant enrichment source. Alternatively, if $N_{\star}$ is low, then it is possible that a pure or washed out Population III signature may still be present in this DLA.\\
\begin{figure*}
    \centering
    \includegraphics[width=1.5\columnwidth]{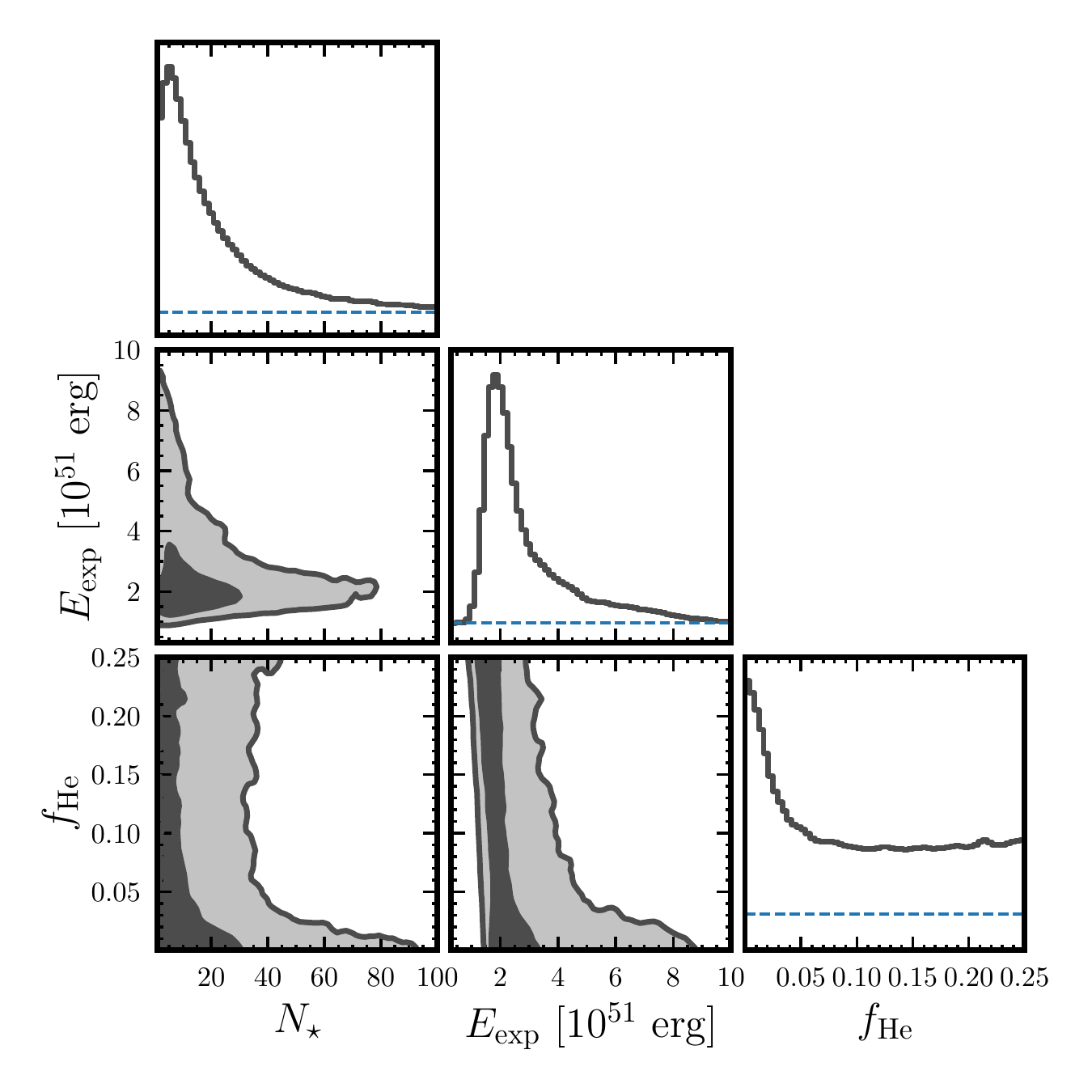}
    \caption{Results of our MCMC analysis of the chemical enrichment of the DLA towards J1001$+$0343 given our stochastic model. From left to right, we show the number of enriching stars, the explosion energy, and the degree of mixing. The diagonal panels indicate the maximum likelihood posterior distributions of the model parameters while the 2D contours indicate the correlation between these parameters. The dark and light gray shaded regions indicate the 68 and 95 per cent confidence intervals, respectively. In the diagonal panels, the horizontal blue dashed line indicates the zero-level of each distribution. 
    }
    \label{fig:corner_stoch}
\end{figure*}
\begin{figure*}
    \centering
    \includegraphics[width=1.5\columnwidth]{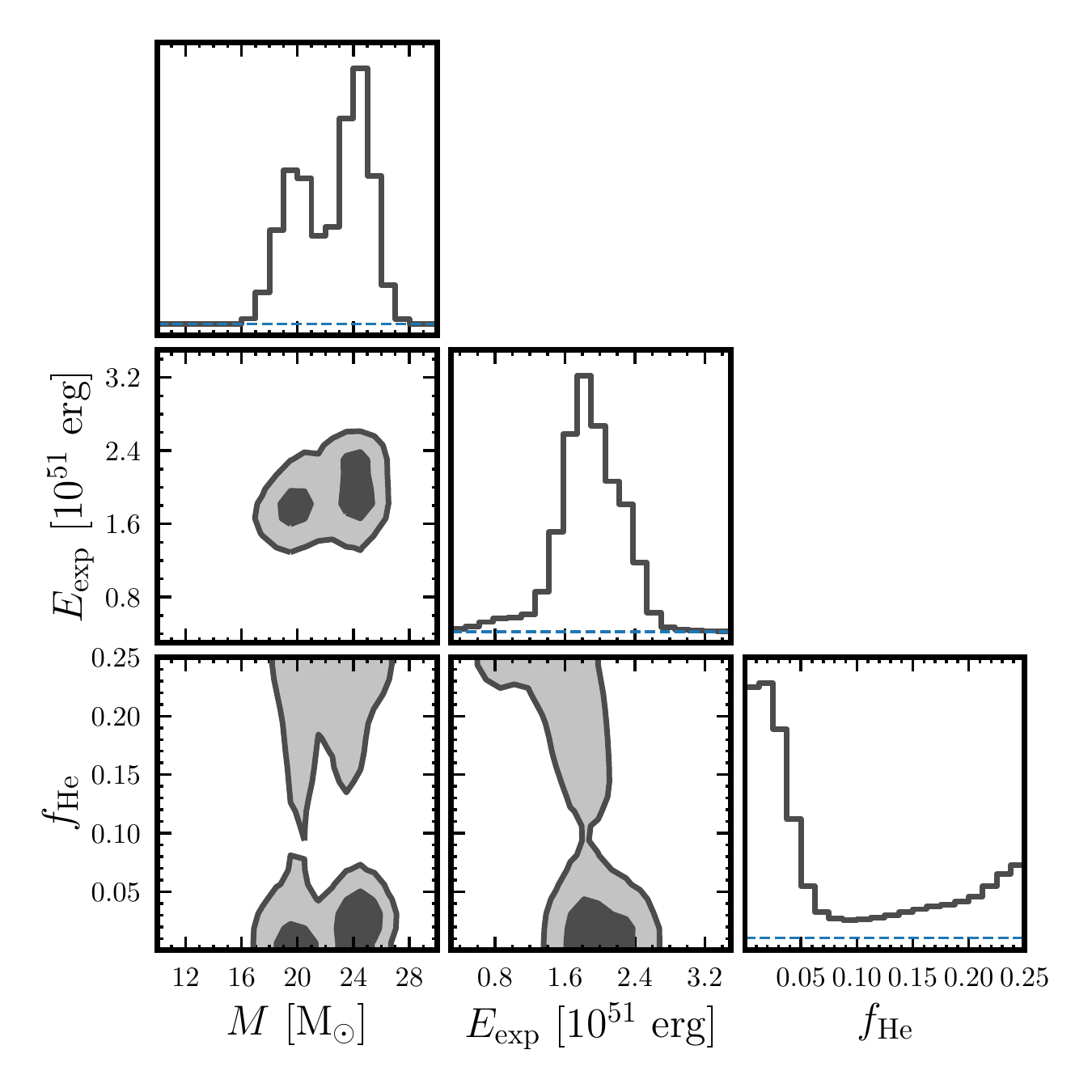}
    \caption{Results of our MCMC analysis of the chemical enrichment of the DLA towards J1001$+$0343 assuming the number of enriching stars $N_{\star}=1$. From left to right, we show the progenitor star mass, the explosion energy, and the degree of mixing. The diagonal panels indicate the maximum likelihood posterior distributions of the model parameters while the 2D contours indicate the correlation between these parameters. The dark and light gray shaded regions indicate the 68 and 95 per cent confidence intervals, respectively. In the diagonal panels, the horizontal blue dashed line indicates the zero-level of each distribution. 
    }
    \label{fig:corner_sing}
\end{figure*}
 \indent We test this idea by using a Markov Chain Monte Carlo (MCMC) likelihood analysis to investigate the number of stars that have enriched this DLA. We explore the entire enrichment model parameter space to find the parameters that best fit our data. The \citetalias{HegerWoosley2010} parameters span $(10-100)~{\rm M}_{\odot}$ (with a mass resolution of $\Delta M\gtrsim0.1~{\rm M}_{\odot}$), $(0.3-10)\times10^{51}$~erg (sampled by 10 values), and $(0-0.25)~f_{\rm He}$ (sampled by 14 values) where $f_{\rm He}$ is the fraction of the He core size; in total, there are 16\,800 models in this yield suite. We adopt a upper mass limit of 70~M$_{\odot}$ beyond which pulsational pair instability SNe are thought to occur \citep{Woosley2017}. This leaves a grid of 15\,792 models to explore. During our analysis, we linearly interpolate between this grid of yields while applying uniform priors on each parameter. 
 The results of this analysis are shown in Figure~\ref{fig:corner_stoch}. 
The most favoured result of this model is that the DLA towards J1001+0343 was enriched by a low number of massive stars which, as argued above, make it consistent with  Population III enrichment (but does not rule out Population II yields).\\
\indent Motivated by these findings, we now assess the possible properties of a putative metal-free star that may be responsible for the enrichment of the DLA towards J1001+0343. In this case we assume that the DLA has been enriched by one Population III SN, again utilising the \citetalias{HegerWoosley2010} yields. 
The results of this analysis are shown in Figure~\ref{fig:corner_sing}. We find that the abundances of this DLA are best modelled by a Population III star with a mass between $19-25$~M$_{\odot}~(2\sigma)$ and an explosion energy between $(0.9-2.4)\times10^{51}$~erg~$(2\sigma)$. The degree of mixing between the stellar layers remains unconstrained, but generally favours lower values of the mixing parameter. To test how well this model describes our data, we compare the [X/O] ratios supported by this model to those presented in Table~\ref{tab:j1001}. This comparison is shown in Figure~\ref{fig:PopIIIyields}. The [C/O], [Si/O], and [Fe/O] of this DLA are simultaneously well described by the inferred likelihood model. \\
The Population III star that best models the abundances of the DLA towards J1001+0343 has an explosion energy that is consistent with the value found for a typical metal-poor DLA \citep{Welsh2019}.  
The results of this analysis are also similar to the inferred enrichment of the most metal-poor DLA currently known; \citet{Cooke2017} find that the abundance pattern of the most metal-poor DLA can be well-modelled by a Population III SN with a progenitor mass $M=20.5~{\rm M}_{\odot}$. 
\begin{figure}
    \centering
    \includegraphics[width=\columnwidth]{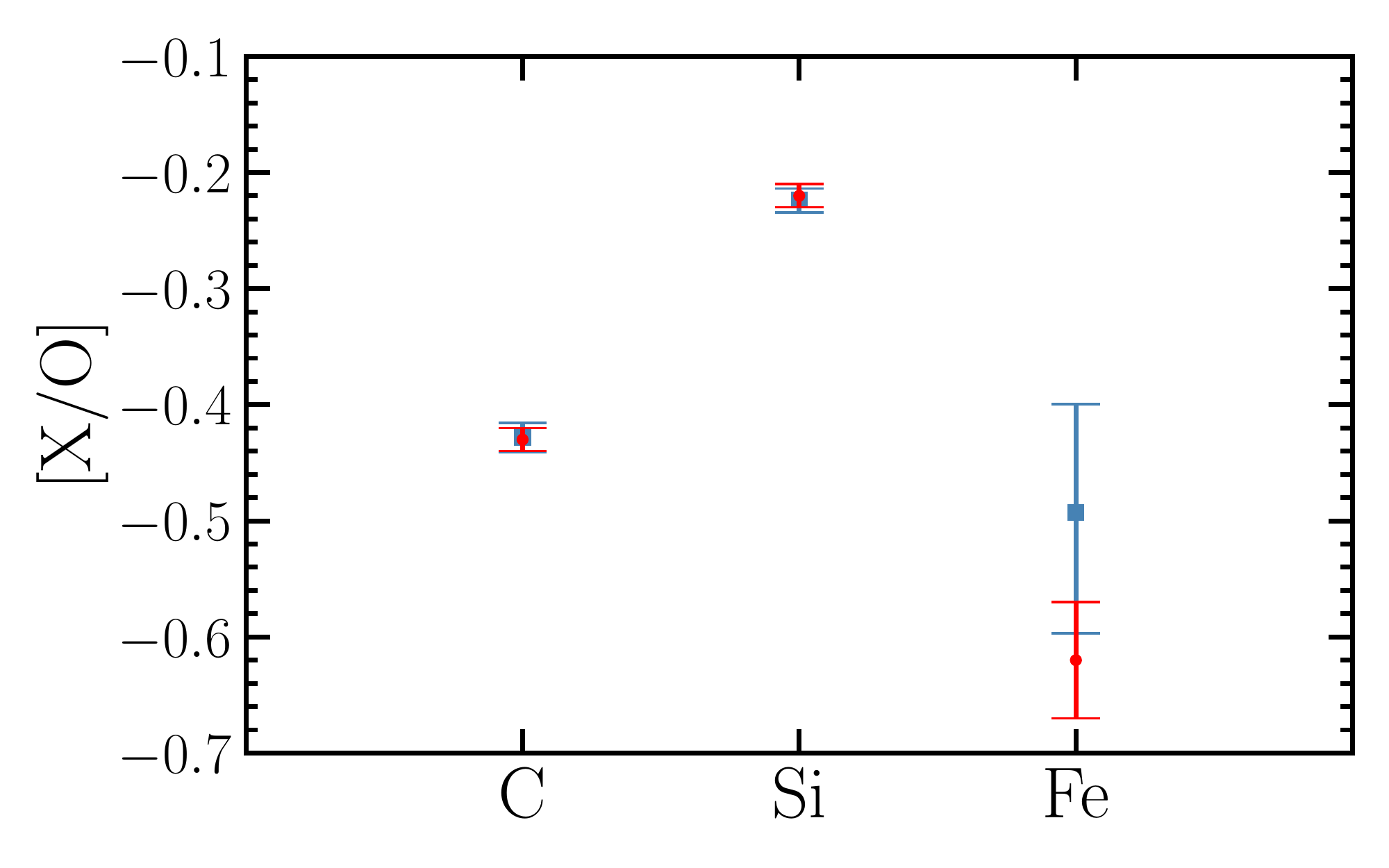}
    \caption{The observed abundances of the DLA towards J1001+0343 (red circles) compared to the best fit abundance ratios inferred by our best fit enrichment model (blue squares). The blue error bars encompass the interquartile range of the model values. }
    \label{fig:PopIIIyields}
\end{figure}
The DLA analysed by \citet{Cooke2017} was preferentially modelled with a somewhat higher explosion energy than that reported here, but still consistent within $2\sigma$. This preference towards higher energy explosions (i.e. hypernovae) is also inferred from the analysis of some EMP stars (e.g. \citealt{GrimmettHeger2018, Ishigaki2018}). However, the \citetalias{HegerWoosley2010}  analysis of the \citet{Cayrel2004} sample of EMP stars favoured models with an explosion energy between $(0.6 - 1.2)\times10^{51}$~erg. Recent theoretical works have started to favour lower energy supernova explosions for metal-free progenitors (e.g. the simulations performed by \citet{Ebinger2020} suggest a range $(0.2 - 1.6)\times10^{51}$~erg). This is in line with the range of explosion energies inferred for the enrichment of the DLA towards J1001+0343. Furthermore, the analysis conducted by \citet{HazeNunez2021} has compared the typical (median) abundances of VMP DLAs to the IMF-weighted yields from various simulations; for the abundances considered in this work,the \citet{Ebinger2020} and \citetalias{HegerWoosley2010}  IMF-weighted yields are equally capable of predicting the typical VMP DLA chemistry. Though, the typical VMP abundances of [C/O], [Si/O], and [Fe/O] favour the \citetalias{HegerWoosley2010} models with higher energy explosions over their lower energy counterparts. \\ 
At present, the models are driven by the relatively low errors on the [C/O] and [Si/O] abundances. Future higher S/N observations and covering a broader wavelength range would allow us to detect additional $\alpha$-capture elements (Mg, S), some odd atomic number elements (e.g. N, Al), and select iron-peak elements (e.g. Ni, Cr, Zn). These elements may allow us to further pin down the properties of the enriching stars. 
An informative probe of the explosion physics is the relative abundance of zinc and iron --- we do not detect zinc absorption in this DLA, but this may be possible with the next generation of $30-40$\,m telescopes.

\subsection{Future avenues for ruling out Population II}
In the previous section we investigated the chemical enrichment of the DLA towards J1001+0343. Under the assumption of a Sapleter IMF, we found that this DLA was best modelled by a low number of enriching stars (consistent with one). We also found that the observed abundances of [C/O], [Si/O], and [Fe/O] can be simultaneously well modelled by the yields of an individual Population III SN. However, given the age of the Universe at redshift $z=3$ ($\sim2$~Gyr), there is  sufficient time ($\sim1.5$~Gyr between $z=10$ to $z=3$) for this DLA to be enriched by Population III stars and subsequent Population II stars. Any putative Population III signature is thought to be washed out soon after the birth of Population II stars \citep{Ji2015}; just a few massive Population II stars are required to wash out a peculiar Population III chemical signature. However, there could be a delay between Population III and Population II star formation. For example, reionization quenching can temporarily suspend star formation in low mass galaxies \citep{Onorbe2015,Wheeler2015}, and this may prolong the time that a Population III signature can be preserved in near-pristine gas. After a period of dormancy, interactions with gaseous streams in the intergalactic medium can help re-ignite star formation in these low mass objects \citep{Wright2019}. Interestingly, the chemistry of the most metal-poor DLAs shows an increase in [C/O] with decreasing redshift \citep[see Figure~8 of][]{Welsh2020}. One interpretation of this trend is that EMP DLAs universally experienced some degree of reionization quenching; the increase in [C/O] with decreasing redshift is interpreted as the onset of enrichment from the carbon yield of the first (i.e. Population II) intermediate mass stars.\\ 
To reduce the possibility of Population II contamination, one might consider measurements of the [O/Fe] ratio close to the epoch of reionization. For example,
\citet{Banados2019} recently reported the detection of a near-pristine DLA at $z=6.4$. While the current determinations of the metallicity and [O/Fe] abundance of this system ([Fe/H]~$=-2.94\pm0.26$; [O/Fe]~$=+0.02\pm0.21$) are too uncertain to study the nature of [O/Fe] in the EMP regime, future higher spectral resolution and higher signal-to-noise ratio observations would allow the chemistry of this gas cloud to be inferred from weak absorption lines, potentially improving the precision of this measurement by an order of magnitude. With the current data, \citet{Banados2019} find no evidence of a Population III chemical fingerprint. The best-fit value of the [O/Fe] value of the \citet{Banados2019} DLA places this system below the typical [O/Fe] abundance of VMP DLAs (at $\sim1.7\sigma$ confidence). 
Considering the high [O/Fe] ratio reported in this paper, if the [O/Fe] value of the \citet{Banados2019} DLA remains unchanged with future observations, these distinct abundances would be a signpost of stochastic chemical enrichment at the lowest metallicity --- this may also be a signature of Population III star formation.\\
\indent While it may be difficult to distinguish the absolute chemical yields of Population II versus Population III stars, it may be possible to identify a transition of Population III to Population II enrichment by empirically looking for a change in the behaviour of chemical abundance ratios as a function of either iron paucity or redshift. Given the tight [O/Fe] plateau seen in VMP DLAs combined with the evidence of an increased [O/Fe] ratio in some EMP DLAs, we thus propose that the [O/Fe] abundance may offer the cleanest probe of the Population III to Population II transition in the most metal-poor DLAs.\\
\subsection{Unravelling the Population III fingerprint}
We have mentioned both EMP stars and EMP DLAs as potential environments to uncover the Population III fingerprint. Both stars and DLAs have their respective advantages. 
The large sample size afforded when studying stellar relics cannot yet be matched in similar studies of gaseous relics. 
The potential evolution of [O/Fe] across EMP stars has also been the subject of much discussion \citep[see the review by][]{McWilliam1997}. However, determining this ratio in EMP stars is particularly challenging. 
There are four approaches to determine the oxygen abundance in stellar atmospheres. The typical method utilises the \oi{777}\,nm triplet; a transition that requires large non-LTE corrections \citep{Fabbian2009}. 
The weak \oi{630}\,nm line is known to form in LTE and, as such, may be more reliable \citep{Asplund2005}. However, the strength of this feature means that it is challenging to detect at low metallicities. Observations of the UV and IR molecular OH features also show systematic offsets relative to the aforementioned [O/H] abundance indicators, although the offsets can be reduced by accounting for 3D hydrodynamical effects \citep{Nissen2002,Garcia2006,Dobrovolskas2015}. Given the difficulty of accurately determining the O abundance in the lowest metallicity stars, we suggest that DLAs are the ideal environment to study the evolution of this element despite the smaller sample size. \\
\indent The comparative analysis of the chemistry of both gaseous and stellar relics offers the opportunity to study early chemical enrichment further. It is well established that some of the most metal-poor Milky Way halo stars show an enhanced [C/Fe] ratio \citep{Beers2005}. This may be a sign of enrichment from the first stars. Or, indeed in some cases, this enhancement may be explained via mass transfer across stars in binary systems \citep{Arentsen2019}. Whether this enhancement is also prevalent across metal-poor DLAs is yet to be seen. Figure~\ref{fig:xfe} highlights that, with current statistics, we cannot discern any concurrent enhancement of the [C/Fe] or [Si/Fe] abundance ratios along with [O/Fe]. The enhancement of [O/Fe] in the EMP regime may not extend to other alpha elements across these DLAs. The lack of concurrent [C/Fe] enhancement may suggest that the CEMP-no stars and the [O/Fe] enhanced DLAs have experienced divergent enrichment histories. This discrepancy may help reveal the origins of CEMP-no stars. \\
\indent Another environment that may offer unique insight to the metals produced by the first stars are the ultra-faint dwarf galaxies (UFDs) that orbit the Milky Way. These UFDs contain some of the most metal-poor stars currently known. 
The stars in these UFDs may have experienced a different enrichment history to the Milky Way halo stars. If metal-poor DLAs are indeed the antecedents of the UFDs, we might be able to search for a consistent chemical signature in the most metal-poor stars of the UFDs.

\section{Conclusions}
\label{ofe:sec:conc}
Previous observations of metal-poor DLAs tentatively suggested that the \emph{most} metal-poor DLAs display an elevated [O/Fe] ratio compared to their higher metallicity counterparts. The higher metallicity ([Fe/H]~$>-3$) DLAs are well described by a plateau around a typical value of $\rm [\langle O/Fe\rangle]~=+0.40 \pm 0.08$. The primary goal of this paper is to assess whether [O/Fe] is indeed elevated amongst the most metal-poor DLAs by presenting a detailed chemical abundance analysis of two near-pristine DLAs --- J0955+4116 and J1001+0343, observed with Keck/HIRES and VLT/UVES respectively. Our main conclusions are as follows:
\begin{enumerate}
    \item  We find that the DLA towards J0955+4116 has a neutral hydrogen column density $\log_{10} N({\rm H}$\,{\sc i}$)/{\rm cm}^{-2} = 20.21\pm0.05$ and a relative iron abundance [Fe/H]~$=-2.95\pm 0.10$. This places the gas cloud towards this quasar on the cusp of the EMP regime. 
    The data collected using Keck/HIRES have revealed [O/Fe]~$=+0.50\pm 0.10$. The [O/Fe] abundance of this DLA is therefore consistent with the plateau observed across DLAs with [Fe/H]~$>-3$.
    \item We present new VLT/UVES data of the DLA toward J1001+0343 that cover previously unobserved Fe\,{\sc  ii} features. These data provide a more precise determination of its chemical composition. We measure an iron abundance of [Fe/H]~$=-3.25\pm0.07$ and an oxygen to iron ratio of [O/Fe]~$=+0.62 \pm 0.05$, reducing the error associated with the [O/Fe] determination by a factor of three compared to previous analyses. 
    \item The [O/Fe] ratio of the DLA towards J1001+0343 is significantly ($2.3\sigma$) above the typical value of a VMP DLA. We have considered the abundances of other ions and this analysis suggests that neither dust depletion nor ionization corrections are the source of this elevation. Rather, the elevated value is intrinsic of the DLA.
    \item The main result of this paper is that the typical [O/Fe] abundance of EMP DLAs ($[\langle{\rm O/Fe}\rangle]=+0.67 \pm 0.04$) is significantly ($3\sigma$) above the typical value of a VMP DLA ($[\langle{\rm O/Fe}\rangle]=+0.40 \pm 0.08$).
    \item The origin of this elevation can be explained if VMP DLAs are all enriched by a similar population of stars, drawn from the same IMF. The divergence at the lowest metallicities likely represents enrichment from a distinct population of stars. Indeed, the scatter associated with the O abundance may highlight the stochastic nature of early chemical enrichment. The chemical composition of this DLA can be well-modelled by the yields of a $(19-25)~{\rm M}_{\odot}~(2\sigma)$ Population III SN with a $(0.9-2.4)\times10^{51}$~erg~$(2\sigma)$ explosion. Given that the adopted yields are also representative of Population II SNe yields, we cannot yet rule out contamination from later stellar populations. 
\end{enumerate} 
We suggest that EMP DLAs display an elevated [O/Fe] compared to their higher metallicity counterparts and this is due to their distinct enrichment histories. The elevated [O/Fe] abundance could be an indicator of enrichment by a generation of metal free stars \citep{HegerWoosley2010}. Further data are necessary to determine if the elevated [O/Fe] ratio of J1001+0343 is typical of an EMP DLA. Forthcoming data on other near-pristine DLAs, as part of this programme, will directly answer this question. Furthermore, upcoming surveys \citep[e.g. DESI, WEAVE, and 4MOST][]{Dalton2012, deJong2012, DESI2016, Pieri2016} will provide new EMP DLA candidates to investigate and improve the statistical significance of this behaviour. If an elevated [O/Fe] abundance can be attributed to enrichment by metal-free stars, then the DLAs analysed in this work may provide a signpost to some of the most pristine environments in the high redshift universe, and would be an ideal place to search for Population III host galaxies and perhaps even the light from Population III SNe using the forthcoming \emph{James Webb Space Telescope}. 

\begin{acknowledgments}
We thank the anonymous referee who provided a prompt and careful review of our paper. We thank M. Fossati for the use of their line-fitting software and we thank A. Skuladottir for helpful discussions surrounding comparisons with stellar populations. This paper is based on observations collected at the European Organisation for Astronomical Research in the Southern Hemisphere, Chile (VLT program IDs: 083.A-0042(A) and 105.20L3.001), and at the W. M. Keck Observatory which is operated as a scientific partnership among the California Institute of Technology, the University of California and the National Aeronautics and Space Administration. The Observatory was made possible by the generous financial support of the W. M. Keck Foundation. The authors wish to recognize and acknowledge the very significant cultural role and reverence that the summit of Maunakea has always had within the indigenous Hawaiian community. We are most fortunate to have the opportunity to conduct observations from this mountain. We are also grateful to the staff astronomers at the VLT and Keck Observatory for their assistance with the observations. 
This work has been supported by Fondazione Cariplo, grant No 2018-2329. 
During this work, R.~J.~C. was supported by a
Royal Society University Research Fellowship.
We acknowledge support from STFC (ST/L00075X/1, ST/P000541/1).
This project has received funding from the European Research Council 
(ERC) under the European Union's Horizon 2020 research and innovation 
programme (grant agreement No 757535).
This work used the DiRAC Data Centric system at Durham University,
operated by the Institute for Computational Cosmology on behalf of the
STFC DiRAC HPC Facility (www.dirac.ac.uk). This equipment was funded
by BIS National E-infrastructure capital grant ST/K00042X/1, STFC capital
grant ST/H008519/1, and STFC DiRAC Operations grant ST/K003267/1
and Durham University. DiRAC is part of the National E-Infrastructure.
This research has made use of NASA's Astrophysics Data System.
\end{acknowledgments}

\facilities{Keck:I (HIRES), VLT: (UVES)}


\vspace{5mm}


\software{Astropy \citep{ASTROPY}, 
Corner \citep{CORNER},
Matplotlib \citep{MATPLOTLIB},
and NumPy \citep{NUMPY}. }



\appendix

\section{The oxygen and iron abundances of our DLA sample}
\label{ofe:append:data}

In this appendix, we present the DLA data used to produce Figure~\ref{fig:ofe_stars}. In Table~\ref{tab:ofe_data} we list the column density of neutral hydrogen, the redshift, and the relative abundances of oxygen and iron.

\begin{table*}
    \centering
        \caption{DLA data used in Figure~\ref{fig:ofe_stars}}
    \label{tab:ofe_data}
    	\tabcolsep=0.65cm
    \begin{tabular}{lccccc}
\hline
QSO & $z_{\rm abs}$ & $\log_{10}N($H\,{\sc i}) & [Fe/H] & [O/Fe] & References \\
\hline \hline
Q0000$-$2620   & 3.390 & 21.41 $\pm$ 0.08 & $-$2.01 $\pm$ 0.09 & +0.33 $\pm$ 0.10 & 1  \\
J0034+1639	 & 4.251 & 20.60 $\pm$ 0.10 & $-$2.84 $\pm$ 0.12 & +0.33 $\pm$ 0.08 & 2  \\
J0035$-$0819  & 2.340 & 20.43 $\pm$ 0.04 & $-$2.89 $\pm$ 0.05 & +0.44 $\pm$ 0.06 &  3, 4, 5\\
HS0105+1619  & 2.537 & 19.42 $\pm$ 0.01 & $-$1.98 $\pm$ 0.09 & +0.22 $\pm$ 0.09 &  6 \\
Q0112$-$306   & 2.418 & 20.50 $\pm$ 0.08 & $-$2.64 $\pm$ 0.09 & +0.40 $\pm$ 0.09 &  7 \\
J0140$-$0839   & 3.697 & 20.75 $\pm$ 0.15 & $-$3.45 $\pm$ 0.24 & +0.70 $\pm$ 0.19 & 3, 8  \\
J0307$-$4945   & 4.467 & 20.67 $\pm$ 0.09 & $-$1.93 $\pm$ 0.19 & +0.48 $\pm$ 0.24 & 9  \\
J0311$-$1722   & 3.734 &	20.30 $\pm$ 0.06  & $<-2.01$ &$>-0.28$ & 3 \\
J0831+3358  & 2.304 & 20.25 $\pm$ 0.15 & $-$2.39 $\pm$ 0.16 & +0.38 $\pm$ 0.08 &  3, 10 \\
J0903+2628  & 3.078 & 20.32 $\pm$ 0.05 & $\leq-$2.81  &  $>-0.24$ &  11 \\
Q0913+072  & 2.618 & 20.34 $\pm$ 0.04 & $-$2.82 $\pm$ 0.04 & +0.42 $\pm$ 0.01 & 12  \\
J0953$-$0504	 & 4.203 & 20.55 $\pm$ 0.10 & $-$2.95 $\pm$ 0.21 & +0.40 $\pm$ 0.19 & 13  \\
J0955+4116  & 3.280 & 20.21 $\pm$ 0.05 & $-$2.95 $\pm$ 0.10 & +0.50 $\pm$ 0.10 &  10, 14 \\
J1001+0343  & 3.078 & 20.20 $\pm$ 0.05 & $-$3.25 $\pm$ 0.07 & +0.62 $\pm$ 0.05 & 3, 14  \\
J1037+0139  & 2.705 & 20.50 $\pm$ 0.08 & $-$2.44 $\pm$ 0.08 & +0.31 $\pm$ 0.04 &  3 \\
Q1108$-$077   & 3.608 & 20.37 $\pm$ 0.07 & $-$1.96 $\pm$ 0.07 & +0.27 $\pm$ 0.04 & 7  \\
J1111+1332  & 2.271 & 20.39 $\pm$ 0.04 & $-$2.27 $\pm$ 0.04 & +0.35 $\pm$ 0.07 & 15  \\
Q1202+3235  & 4.977 & 19.83 $\pm$ 0.10 & $-$2.44 $\pm$ 0.16 & +0.42 $\pm$ 0.14 &  16 \\
J1340+1106  & 2.508 & 20.09 $\pm$ 0.05 & $-$2.07 $\pm$ 0.05 & +0.31 $\pm$ 0.04 &  3 \\
J1340+1106  & 2.796 & 21.00 $\pm$ 0.06 & $-$2.15 $\pm$ 0.06 & +0.50 $\pm$ 0.04 &  3 \\
J1358+0349  & 2.853 & 20.16 $\pm$ 0.02 & $-$3.32 $\pm$ 0.18 & +0.70 $\pm$ 0.18 &  17 \\
J1358+6522  & 3.067 & 20.50 $\pm$ 0.08 & $-$2.88 $\pm$ 0.08 & +0.54 $\pm$ 0.03 &  17 \\
J1419+0829  & 3.050 & 20.40 $\pm$ 0.03 & $-$2.33 $\pm$ 0.04 & +0.41 $\pm$ 0.04 &  3 \\
J1558$-$0031  & 2.703 & 20.67 $\pm$ 0.05 & $-$1.95 $\pm$ 0.05	& +0.38	$\pm$ 0.04 & 18  \\
J1558+4053  & 2.553 & 20.30 $\pm$ 0.04 & $-$2.70 $\pm$ 0.07 & +0.25 $\pm$ 0.07 &  12 \\
Q1946+7658   & 2.844 & 20.27 $\pm$ 0.06 & $-$2.50 $\pm$ 0.06 & +0.36 $\pm$ 0.01 & 19  \\
Q2059$-$360   & 3.083 & 20.98 $\pm$ 0.08 & $-$1.97 $\pm$ 0.08 & +0.39 $\pm$ 0.04 &  7 \\
J2155+1358   & 4.212 & 19.61 $\pm$ 0.10 & $-$2.15 $\pm$ 0.25 & +0.35 $\pm$ 0.24 & 20  \\
Q2206$-$199  & 2.076 & 20.43 $\pm$ 0.04 & $-$2.57 $\pm$ 0.04 & +0.50 $\pm$ 0.03 &  12 \\
\hline
    \end{tabular}
    \begin{tabular}{l}
      1) \citet{Molaro2000}; 2) \citet{Berg2016}; 3) \citet{Cooke2011b};  4) \citet{Cooke2011a};   5) \citet{Welsh2020}; \\
      6) \citet{Omeara2001};  7) \citet{Petitjean2008}, 8) \citet{Ellison2010};   9) \citet{Dessauges-Zavadsky2001}; \\
  10) \citet{Penprase2010};  11) \citet{Cooke2017};  12) \citet{Pettini2008};  
  13) \citet{Dutta2014};  14) This work; \\
      15) \citet{CookePettiniJorgenson2015};    16) \citet{Morrison2016};  17) \citet{Cooke2016};
      18) \citet{OMeara2006}; \\ 19) \citet{Prochaska2002b}; 
      20) \citet{Dessauges-Zavadsky2003}. 
\end{tabular}

\end{table*}

\newpage
\bibliography{references}{}
\bibliographystyle{aasjournal}



\end{document}